\documentstyle[11pt]{article}

\newcommand{\dddot}[1]{\stackrel{...}{#1}}
\newcommand{\ddddot}[1]{\stackrel{....}{#1}}
\newcommand{\be}{\begin{eqnarray*}}
\newcommand{\ee}{\end{eqnarray*}}
\newcommand{\bd}{\begin{description}}
\newcommand{\ed}{\end{description}}
\newcommand{\bi}{\begin{itemize}}
\newcommand{\ei}{\end{itemize}}
\newcommand{\bc}{\begin{center}}
\newcommand{\ec}{\end{center}}

\newcommand{\nn}{\nonumber}

\input amssym.def

\begin{document}

\title{ Newtonian and Post-Newtonian approximations of the $k = 0$ Friedmann Robertson Walker Cosmology.}
\author{Tamath Rainsford\\
Department of Physics and Mathematical Physics, \\
University of Adelaide, South Australia 5005, Australia}
\date{}
\maketitle

\begin{abstract}
In a previous paper \cite{szek-rain}, we derived a
post-Newtonian approximation to cosmology which, in contrast to former
Newtonian and post-Newtonian cosmological theories, has a well-posed initial
value problem. In this
paper, this new post-Newtonian theory is compared with the fully general
relativistic theory, in the context of the $k = 0$ Friedmann Robertson Walker
cosmologies. It is found that the post-Newtonian theory reproduces the results of
its general relativistic counterpart, whilst the Newtonian theory
does not.  
\end{abstract}

\section{Introduction}

The equations of general relativity are
difficult to solve and often become tractable only in a Newtonian
context. Hence, it is desirable to use Newtonian theory rather than general relativity
 where possible \cite{ellis}. By Newtonian, we mean that theory which
is  obtainable from general relativity by taking the limit of weak gravitational
fields and small velocities. Unfortunately, when Newtonian theory is
applied to cosmology, the boundary conditions become redundant, and subsequently the Poisson
equation  no longer has a unique solution. Thus, the theory is no longer
well-posed and causality may be violated.

In a previous work \cite{szek-rain}, we explored the insufficiencies of the
Newtonian theory in detail and showed that in addition to the lack of
well-posedness Newtonian theory is also incomplete. The reasons for these
insufficiencies are the following. In Newtonian theory, which is the
expansion of general relativity in terms of weak gravitational fields and
small velocities up to order $c^{-2}$, all that remains of the field
equations is the Poisson equation, ie. the evolution equation and the constraint equations have become one and the same. With no
time evolution equation for the Newtonian potential $\phi$, a unique solution
for $\phi$ cannot be found, and so the theory is not well-posed. Furthermore,
the term $-\bigtriangledown \phi$, a term of order $c^{-4}$,  does not feature in the
field equations, and hence the Bianchi identities cannot be fully
obtained. Although completeness  is obtainable at order $c^{-4}$, it is
necessary to go to order $c^{-6}$ and then to 
reformulate the field equations as wavelike equations, in order to obtain a
well-posed initial value problem. Throughout the remainder of this paper we
will call this approximation to general relativity ``post-Newtonian
theory''.

In the following we will explore the Friedmann Robertson Walker (FRW) cosmology in the
context of the above Newtonian and post-Newtonian approximations. It has been shown
that the FRW models have a simple Newtonian
interpretation \cite{m-r, bondi}. Howsoever, it will be outlined here that
the Newtonian theory cannot reproduce
all the solutions of the fully general relativistic theory: Variations in the
equation of state do not cause changes in the solutions of the theory since
the pressure does not enter into the dynamics. This is
reflected in the Raychaudhuri equation being reproducible only for the
case of vanishing pressure. Thus, the Newtonian
theory is only useful for the special case of dust. In the post-Newtonian
theory, on the other hand, 
such difficulties are overcome. The pressure does feature in the dynamics,
and varying the equation of state produces correspondingly varying solutions,
allowing for the full range of possibilities of its general relativistic
counterpart. We therefore argue that the post-Newtonian theory should be used
whenever we have non-zero pressure in the univese.

In Section 2 we write down what constitutes the general
relativistic $k = 0$ FRW cosmology. In Section \ref{NandP} we will introduce the Newtonian and post-Newtonian
approximations and  put the FRW metric into a form
such that we may draw comparisons between these three theories. In Section \ref{n} we explore the 
Newtonian approximation, and we will show that the theory may only be used in the
special case of dust. In Section \ref{pn} we will find that the post-Newtonian theory is
able to fully reproduce the results of the general relativistic case. We will
discuss our results in Section \ref{conc}.

\section{The Fully General Relativistic $k = 0$ Friedmann Robertson Walker
Cosmology}
\label{fgr}

In the case of the flat FRW metric, the field
equations are the Friedmann equation and the Raychaudhuri equation 

\begin{eqnarray}
\left(\frac{\dot{R}}{R}\right)^2  =  \frac{8}{3}\pi G \rho, \label{Friedmann} 
\end{eqnarray}
and
\begin{eqnarray}
3\frac{\ddot{R}}{R} = - 4\pi G (\rho + 3pc^{-2}),
\label{Ray}
\end{eqnarray}
from which one may obtain the Bianchi identity
\begin{eqnarray}
\dot{\rho} + 3(\rho + pc^{-2})\frac{\dot{R}}{R} = 0.\nn
\end{eqnarray}

Assuming an equation of state of the form $p = w \rho c^2$ implies $\rho = \frac{3C}{8\pi G}
R^{-3(1 + w)}$, with $C$ a constant. Thus, from (\ref{Friedmann}) it follows that 

\begin{eqnarray}
R(t) = \left(\frac{3}{2}(1+w)C^{\frac{1}{2}}t\right)^{\frac{2}{3(1+w)}}. \nn
\end{eqnarray}

Typical values for $w$ include: \\

\begin{eqnarray}
{\rm Matter\:\:\:(w = 0):  \hspace{1.5cm} \rho \propto R^{-3}, \hspace{1.5cm} R \propto
t^{\frac{2}{3}},} \nn\\[1cm]
{\rm Radiation\:(w = \frac{1}{3}):  \hspace{1.5cm} \rho \propto R^{-4},
\hspace{1.5cm} R \propto t^{\frac{1}{2}},} \nn\\[1cm]
{\rm Stiff\:Matter\:(w = 1):  \hspace{1.5cm} \rho \propto R^{-6}, \hspace{1.5cm}
R \propto
t^{\frac{1}{3}}.} 
\label{sol} 
\end{eqnarray}

\section{Newtonian and Post-Newtonian Approximations and the Friedmann
Robertson Walker Metric}
\label{NandP}

We now would like to study the FRW cosmology in the
Newtonian and post-Newtonian approximations. Following a scheme similar to
that of 
Weinberg's \cite{wein}, we adopt units in which the typical velocity has
magnitude 1, i.e.\ $\beta \approx c^{-1}$, and assume a one parameter family
of metrics $g_{\mu\nu}(x^{\lambda},c)$ for which there is a system of coordinates
$(x^0, x^i)$ in which the components of the metric have the following asymptotic behaviour as $c \longrightarrow \infty$:

\begin{eqnarray}
g_{00} & = & -1 -2 \phi c^{-2} - 2 \alpha c^{-4} - 2 \alpha' c^{-6} - 2
\alpha''c^{-8}..... \;, \nonumber \\
g_{0i} & = & \zeta_i c^{-3} + \zeta_i' c^{-5} + \zeta_i'' c^{-7}.....\;, \nn \\
g_{ij} & = & \delta_{ij} - 2 \phi \delta_{ij} c^{-2} + \alpha_{ij} c^{-4} 
                + \alpha_{ij}' c^{-6} + \alpha_{ij}'' c^{-8}.....\;.\label{pnform}
\end{eqnarray}

The usual Newtonian theory is obtained as the ${\cal O}(c^{-2})$ limit of
(\ref{pnform}), while the complete Newtonian approximation is the ${\cal
O}(c^{-4})$ limit. Reformulating the field equations of the ${\cal O}(c^{-6})$ limit as
wavelike equations defines the post-Newtonian theory (see \cite{szek-rain}).

We need to put the $k = 0$ FRW metric, \\

\hspace{2.5cm} $ds^2 = -dx_0^2 + \Sigma_i R(t)^2 dx_i^2 \label{FLRW},$  \hspace{2cm} where $x_0 = ct$, \\
into a form of which we
can read off the potentials $\phi$, $\zeta_i$, $\alpha$ and $\alpha_{ij}$.  To do so, we consider the following coordinate transformation

\begin{eqnarray}
x_0 & = & Tc + \tau c^{-1} + \tau' c^{-3},\nn \\
x_i & = & R^{-1} X_i + \chi_i c^{-2} + \chi'_i c^{-4}, \nn 
\end{eqnarray}
with

\begin{eqnarray}
\tau & = & A(t) + A_{ij}(t)X_{ij}, \nn \\
\tau' & = & B_{ij}(t)X_{ij} + B_{ijkl}(t)X_{ijkl}, \nn\\
\chi_{i} & = & C_{ij}(t)X_j + C_{ijkl}(t)X_{jkl}, \nn \\
\chi'_i & = & D_{ijkl}(t)X_{jkl} + D_{ijklmn}(t)X_{jklmn}, \nn
\end{eqnarray}
where $X_{ij} \equiv X_iX_j$ and similar for $X_{ijk}$ etc. Throughout the
remainder of the paper we assume: $\dot{A} \ll c^{2}$ to ensure
convergence of our expansion in $c^{-2}$. $ A(t)$, $A_{ij}(t)$, $B_{ij}(t)$,
$B_{ijkl}(t)$, $C_{ij}(t)$, $C_{ijkl}(t)$, $D_{ijkl}(t)$ and $D_{ijklmn}(t)$
are arbitrary functions of time. In these coordinates the metric becomes

\begin{eqnarray}
ds^2 & = & c^2 dT^2 \left[ -1 + c^{-2} (-2\dot{A} - 2\dot{A}_{ij}X_{ij} +
\left(\frac{\dot{R}}{R}\right)^2 X_{ii})\right. \nn \\
&& + c^{-4} (-2\dot{B}_{ij}X_{ij} - 2\dot{B}_{ijkl}X_{ijkl} + (\dot{A} +
\dot{A}_{ij}X_{ij})^2 - 2\dot{R}\dot{C}_{ij}X_{ij} -
2\dot{R}\dot{C}_{ijkl}X_{ijkl} \nn \\ 
&& \left. - 2\left(\frac{\dot{R}}{R}\right)^2\dot{A}X_{ii} -
2\left(\frac{\dot{R}}{R}\right)^2\dot{A}_{ij}X_{kk}X_{ij}) + {\cal O}
(c^{-6})\right] \nn \\
&& + cdT dX_i \left[c^{-1}(-4A_{ij}X_j - 2\frac{\dot{R}}{R}X_i) \right. \nn \\
&& + c^{-3}(- 4B_{ij}X_j - 8 B_{ijkl}X_{jkl} - 4\left(\frac{\dot{R}}{R}\right)^2A_{ij} X_{kk}X_j +
2R\dot{C}_{ij}X_j + 2R\dot{C}_{ijkl}X_{jkl} \nn \\
&& \left. - 2\dot{R}C_{ij}X_j - 6\dot{R}C_{ijkl}X_{jkl} + 2\frac{\dot{R}}{R} \dot{A}X_i + 2 \frac{\dot{R}}{R}
\dot{A}_{jk} X_{ijk}) + {\cal O}(c^{-5})\right] \nn \\
&& + dX_idX_j \left[ \delta_{ij} + c^{-2}(-4 A_{ik}A_{jl}X_{kl} + 2RC_{ij} +
6RC_{ijkl}X_{kl} + 4 \frac{\dot{R}}{R}A_{ik}X_{jk})\right. \nn \\
&& + c^{-4}(-8\dot{A}A_{ik}A_{jl}X_{kl}  - 8B_{ik}A_{jl}X_{kl} +
8A_{ik}A_{jl}\dot{A}_{mn}X_{klmn} - 16 A_{jl}B_{ikmn}X_{klmn} + 6 R
D_{ijkl}X_{kl} \nn\\
&& + 10 RD_{ijklmn}X_{klmn} + R^2(C_{ij} + 3C_{ijkl}X_{kl})^2 + 4
\left(\frac{\dot{R}}{R}\right)^2A_{ik}A_{jl}X_{kl}X_{mm} + 4\dot{R}A_{il}C_{jk}X_{lk}
\nn\\
&& - 4RA_{in}\dot{C}_{jmkl}X_{mnkl} - 4R A_{il}\dot{C}_{jk}X_{lk} +
12\dot{R}A_{in}C_{jmkl}X_{mnkl} - 4\frac{\dot{R}}{R} \dot{A}A_{il}X_{jl} \nn
\\
&& \left. - 4 \frac{\dot{R}}{R}A_{im}\dot{A}_{kl}X_{jklm} + 4
\frac{\dot{R}}{R}B_{ik}X_{jk} + 8 \frac{\dot{R}}{R}B_{imkl}X_{mklj}) + {\cal
O}(c^{-6})\right]. \label{metric}
\end{eqnarray}

Since in (\ref{pnform}) there are no terms of order $c^{-1}$, 
we require that

\begin{eqnarray}
A_{ij} = - \frac{1}{2} \frac{\dot{R}}{R} \delta_{ij}. \label{A!}
\end{eqnarray}

In addition, the terms of order $c^{-2}$ in $g_{00}$ and $g_{ij}$ may be
identified, and from the powers in $X_i$ one can read off

\begin{eqnarray}
C_{ij} = - \frac{\dot{A}}{R} \delta_{ij}, \nn
\end{eqnarray}
and 

\begin{eqnarray}
C_{ijkl} = \frac{1}{18R} \left( \frac{\ddot{R}}{R} - \left(\frac{\dot{R}}{R}\right)^2\right)\delta_{i(j}\delta_{kl)}. \nn
\end{eqnarray}

\section{The Homogeneous and Isotropic Newtonian Cosmological Theory}
\label{n}

The Newtonian theory is the $c^{-2}$ cut-off of (\ref{pnform}) and consists of
the field (Poisson) equation 

\begin{eqnarray}
\phi_{,ii} = 4\pi G\rho,
\label{Poisson}
\end{eqnarray}

($\phi$ is the Newtonian potential and $\rho$ is the density) and the
continuity and Euler equations of fluid dynamics

\begin{eqnarray}
\dot{\rho} + \rho v_{i,i} = 0, \label{Continuity} \\
\dot{v_i} + \phi_{,i} + \frac{1}{\rho}p_{,i} = 0, \label{Euler}
\end{eqnarray}
where $v_i$ is the velocity field and $p$ is the pressure. Homogeneity implies that the density and pressure are merely functions of
time and that the velocity field is  the same relative to all observers. It
can be shown, \cite{heck-shuck} and \cite{szek-rank}, that this amounts to $v_i = V_{ij}(t)  X_j$. By
substituting the Poisson equation into the Euler equation we see that the
Newtonian potential must be of the form $\phi = a_{ij}(t)X_iX_j + a(t)$. Therefore, the
Newtonian approximation of a homogeneous cosmology is

\begin{eqnarray}
&& a_{ii} = 4 \pi G \rho, \nn \\
&& \dot{\rho} + \rho V_{ii} = 0, \nn \\
&& \dot{V_{ij}} + V_{ik}V_{kj} = a_{ij}. \label{nc}
\end{eqnarray}

The FRW cosmology is the most general isotropic and
homogeneous solution. Thus, we will only consider the case where (\ref{nc})
becomes isotropic, ie. shear-free and rotation-free. Heckmann and
Sch\"{u}cking in \cite{heck-shuck2} formulate the more general anisotropic
Newtonian cosmology where there is shear and rotation. To this end we make the following decomposition 

\begin{eqnarray}
V_{ij} = \frac{1}{3} \theta \delta_{ij} + \sigma_{ij} + w_{ij}, \label{decomposition}
\end{eqnarray}

where  

\begin{eqnarray}
\theta & = & V_{ii}, \nn \\
\sigma_{ij} & =  & \frac{1}{2} (V_{ij} + V_{ji}) - \frac{1}{3} \theta \delta_{ij}, \nn \\
w_{ij} & = & \epsilon_{jik} w_k = \frac{1}{2}(V_{ij} - V_{ji}), \nn
\end{eqnarray}

The trace part $\theta$ is the expansion, the tracefree symmetric piece
$\sigma_{ij}$ is the shear and $w_{ij}$, the anti-symmetric part,
is the rotation. Using this decomposition in the Euler equation and setting
$\sigma = 0$ and $w = 0$, we get 

\begin{eqnarray}
\dot{\theta} =  - \frac{1}{3}\theta^2 - 4\pi G\rho, \nn \\
a_{ij} = \frac{1}{3}a_{kk}\delta_{ij}, \nn
\end{eqnarray}
with continuity equation

\begin{eqnarray}
\dot{\rho} + \rho \theta = 0. \label{cont}
\end{eqnarray}

Defining a function $R'(t)$ such that $\theta = 3\frac{\dot{R'}}{R'}$; then the solution of the continuity equation (\ref{cont}) is 

\begin{eqnarray}
\rho = C' R'^{-3},
\label{rho}
\end{eqnarray}
with $C'$ a constant.  Finally, using (\ref{rho}) in the Euler equation, we see
that the Newtonian, isotropic, homogeneous cosmology is given by 

\begin{eqnarray}
a_{ii} & = & 4\pi G\rho, \label{1} \\
\frac{\ddot{R'}}{R'} & = & - \frac{4}{3} \pi G \rho, \label{2} \\
\rho & = & C'R'^{-3}. \label{3}
\end{eqnarray}

Notice here the use of $R(t)$ as the general relativistic scale factor in
(\ref{Friedmann}) through to (\ref{sol}), and $R'(t)$ as the Newtonian scale factor in
the Newtonian theory (\ref{1}) through to (\ref{3}). We are now in a position to
compare these two theories. The general relativistic theory is well-posed. Equations (\ref{Friedmann}) and (\ref{Ray}) are consistent with the
Bianchi identities. In the Newtonian theory there is only one field equation
(\ref{1}), and there is no completeness because (\ref{1}) does not give (\ref{2}) and
(\ref{3}), and nor is the theory well-posed. Also, notice that
in the general relativistic theory pressure occurs in the dynamics of the
theory, whereas in the Newtonian theory, pressure does not occur anywhere in
the dynamics and is only defined through an equation of state.

Equation (\ref{2}) has the same form as the Raychaudhuri equation (at least
when $p = 0$).  Using the Raychaudhuri equation it may be deduced that $R' \propto t^{\frac{2}{3}}$. Thus the equations (\ref{2}) and (\ref{3}) of the Newtonian theory predict the same results, at least for the case of matter, that the general relativistic equations (\ref{Friedmann}) and (\ref{Ray}) do.

How, if
at all, do $R(t)$ and $R'(t)$ differ? To answer this question we use the
remaining piece of information - (\ref{1}), the Poisson equation. Since $\phi
= a_{ij}(t)X_i X_j + a(t)$ is the term of order $c^{-2}$
in the $c^2dT^2$ piece of the FRW metric (\ref{metric}); a comparison of  
 (\ref{pnform}) and (\ref{metric}), and using (\ref{A!}) yields

\begin{eqnarray}
a_{ii} & = & - 3 \frac{\ddot{R}}{R}, \nn
\end{eqnarray}
and
\begin{eqnarray}
a(t) = \dot{A}, \nn
\end{eqnarray}
from which, with the aid of (\ref{1}), we can deduce  

\begin{eqnarray}
\frac{\ddot{R}}{R} = - \frac{4}{3}\pi G \rho. \label{nop}
\end{eqnarray}

This is again the Raychaudhuri equation of general relativity for the case of
vanishing pressure. (Since the Newtonian potential $\phi$ only
appears as $\phi_{,kk} = 2a_{kk}(t)$ in equations (\ref{Poisson}) to (\ref{Euler}) we may
set the piece $a(t) = 0$ without loss of generality; then $A$ is just a constant.)

Hence, the correct general relativistic scale factor
$R(t)$ is equivalent to the Newtonian scale factor $R'(t)$. Dautcourt
\cite{daut}, shows how the validity of the Friedmann equation within Newtonian
cosmology can be understood: Newtonian cosmology is applicable
only when confined to a neighbourhood of the observer, corresponding to
distances which are small compared to the Hubble distance.

Although reproducing results similar to the general relativistic theory, the
Newtonian theory suffers in that varying the equation of state will have no
effect on the outcome of the solutions for $\rho (t)$ and $R'(t)$. This being due
to the fact that the pressure has not appeared anywhere in
the dynamics.  Thus we can only reproduce the
results of the matter dominated case of general relativity.

\section{The Homogeneous and Isotropic Post-Newtonian Cosmological Theory}
\label{pn}

The field equations for the post-Newtonian theory \cite{szek-rain} are 

\begin{eqnarray}
\phi_{,kk} & = & 4\pi G \rho + \frac{1}{4c^2}(- \phi_{jk,jk} - {\frak A}), \label{C1} \\
\zeta_{i,kk} & = & 16 \pi G\rho v_i  + \frac{1}{c^2} \left(
\dot{\phi}_{ij,j} - {\frak B}_i \right).  \label{C2i} \\
\ddot\phi_{ij} - c^2\phi_{ij,kk} & = &  {\frak B}_{ij} + c^2\left[16\pi G
(\rho v_i v_j + \delta_{ij} p) - {\frak A}_{ij} \right],
\label{N3ij}
\end{eqnarray}
where the $\alpha$ and $\alpha_{ij}$ of (\ref{pnform}) are such that

\begin{eqnarray}
\phi_{ij} = \alpha_{ij} - 2 \delta_{ij} \alpha, \nn
\end{eqnarray}
with

\begin{eqnarray}
{\frak A} & \equiv & 6\phi_{,i}\phi_{,i} - 16 \pi G (\rho v^2 + 4\rho\phi), \nn \\
{\frak B}_i & \equiv &  3\zeta_{j,j}\phi_{,i} +
2\zeta_j\phi_{,ij} - 2\phi_{,j}\zeta_{j,i}\nonumber \\
& & \quad - 16\pi G\left[v_i p + \rho v_iv^2 -
\frac{1}{2}\rho \zeta_i\right], \nn\\ 
{\frak A}_{ij} & \equiv & 8\phi\phi_{,ij} + 4\phi_{,i}\phi_{,j} - \delta_{ij}
(6\phi_{,k}\phi_{,k} + 32\pi G\rho\phi),\nn \\
{\frak B}_{ij} & \equiv & -\frac{1}{2}\left(\zeta_i\zeta_{k,kj} + \zeta_j\zeta_{k,ki}\right) - \zeta_k\left(\zeta_{i,jk} + \zeta_{j,ik}\right) + 2\zeta_k\zeta_{k,ij} + \zeta_{k,i}\zeta_{k,j} + \zeta_{i,k}\zeta_{j,k}\nonumber \\ 
&& - 2\phi_{,k}\left(\phi_{ki,j} + \phi_{kj,i} - 2\phi_{ij,k}\right)
- 16\phi\phi_{,i}\phi_{,j} + \phi_{,i}\phi_{kk,j} \nonumber\\
& & + \phi_{,j}\phi_{kk,i} - 2\phi\left(\phi_{ki,jk}
+\phi_{kj,ik} - \phi_{ij,kk} - \phi_{kk,ij}\right) -
2\phi_{ki}\phi_{,jk} \nonumber\\
& &  - 2\phi_{kj}\phi_{,ik} + 2\phi_{,ij}\phi_{kk}
- \delta_{ij}\left[\frac{1}{2}\zeta_{k,m}\zeta_{k,m} +
\frac{1}{2}\zeta_{m,k}\zeta_{k,m} + \frac{1}{2}(\zeta_{k,k})^2
\right.\nonumber\\
& &  - \zeta_k\zeta_{m,mk} - 4\phi_{,k}\phi_{km,m} +
4\phi_{,k}\phi_{mm,k} - 12\phi\phi_{,k}\phi_{,k} \nonumber\\ 
& & - \phi\left(2\phi_{km,mk} - 2\phi_{mm,kk}\right)] + 8\pi G\left[2pv_iv_j \right.\nonumber\\
& &  + 2\rho(2\phi + v^2)v_iv_j + \rho\phi_{ij}
+\delta_{ij}\left(2\rho \phi v^2 - \frac{1}{2}\phi p\right.\nonumber\\
& & \left.\left. + \frac{3}{4}\phi_{,k}\rho_{,k} + \frac{1}{2}\rho
\phi_{,kk}\right)\right]. \nn 
\end{eqnarray}

Along with the harmonic gauge conditions

\begin{eqnarray}
\dot{\phi} & = & - \frac{1}{4}\zeta_{i,i},\label{hg2}\\
\dot{\zeta_i} & = & \phi_{ij,j}, \label{hg1}
\end{eqnarray}
this system forms a closed set which is consistent because the
Bianchi identities are obtainable from the field equations:

\begin{eqnarray}
\lefteqn{\dot\rho \left(1 + \frac{v^2 - 4\phi}{c^2}\right) + (\rho v_j)_{,j}\left(1 + \frac{v^2}{c^2}\right) + \frac{1}{c^2} \left[
\rho\left(2v_j\dot v_j + 2v_jv_k v_{k,j} +\frac{1}{2}\zeta_{j,j}\right)\right.  }\nonumber\\
& & \left. - \frac{1}{2}\rho_{,j}\zeta_j + (v_jP)_{,j} +
\frac{1}{16 \pi G}(2\phi_{,i}\zeta_{i,jj} - 2\zeta_i\phi_{,jii} 
- 3\zeta_{i,i}\phi_{,jj}) \right] = 0,
\label{B1}
\end{eqnarray}
and 
\begin{eqnarray}
\lefteqn{\rho(\dot{v}_i + v_{i,j}v_j + \phi_{,i}) + P_{,i} = \frac{1}{16\pi G
c^2} \left[ -(\dot{{\frak A}} + {\frak B}_{j,j})v_i + \right. }\nonumber \\
& &  \left. \dot{{\frak B}}_i - {\frak B}_{ij,j} - 2\phi({\frak A}_{,i}
+\phi_{jk,kij}) - \phi_{,i}({\frak A} + \phi_{jk,jk}) \right]. \label{B2i}
\end{eqnarray}

 Specifically,
(\ref{B1}) and the $,i$ derivative of (\ref{C2i}) imply $\partial/\partial t$
of the first constraint equation (\ref{C1}), while (\ref{B2i}) and $,i$ of
eq.\ (\ref{N3ij}) imply the time derivative of (\ref{C2i}).

To make comparisons with the FRW
cosmologies we need to consider the case where this theory is both
homogeneous and isotropic. The following are the most general ans\"atze, providing isotropy and homogeneity, for
tensors of rank n, expanded up to order $X^{n + 2}$:

\begin{eqnarray}
\phi & = & ^{(2)}\phi (t) X^2 + \:^{(1)}\phi (t), \nn \\
\zeta_i & = & ^{(2)}\zeta (t) X^2 X_i + \:^{(1)}\zeta (t) X_i, \nn \\
\phi_{ij} & = & ^{(5)}\tilde{\phi}(t) X^4 \delta_{ij}  + \:^{(4)}\tilde{\phi}
(t) X^2 X_{ij} + \:^{(3)}\tilde{\phi} (t) X^2 \delta_{ij} + \:^{(2)}\tilde{\phi}
(t) X_{ij} +  \:^{(1)}\tilde{\phi} (t) \delta_{ij}. \label{ant}
\end{eqnarray}

Again, we decompose the velocity field, $V_{ij}$, (\ref{decomposition})
 and set $w_{ij} = 0$ and $\sigma_{ij} = 0$. Thus we have $v_i =
 V_{ij}X_j = \frac{1}{3} \theta' X_i$. Pro tempore we do not make the substitution $\theta
 = 3 \frac{\dot{R}'}{R'}$, instead, we put $\theta'$.  The field equations,
 (\ref{C1}) to (\ref{N3ij}), are then

\begin{eqnarray}
&& 3\: ^{(2)}\phi = 2\pi G \rho + c^{-2}\left[ \frac{-3}{4}\: ^{(3)}\tilde{\phi} -
\frac{3}{2}\: ^{(2)}\tilde{\phi} + 8\pi G \rho ^{(1)}\phi \right], \label{F1} \\[1cm]
&& 10\: ^{(5)}\tilde{\phi} + 15\: ^{(4)}\tilde{\phi} + 12\: ^{(2)}\phi^2 - 8\pi
G\rho \left(\frac{1}{9} \theta '^2 + 4\: ^{(2)}\phi \right) = 0, \label{F2} \\[1cm]
&& 5\: ^{(2)}\zeta - 8\pi G\rho \frac{1}{3} \theta ' + c^{-2}\left[-
^{(3)}\dot{\tilde{\phi}} - 2\: ^{(2)}\dot{\tilde{\phi}} + 9\: ^{(2)}\phi \:
^{(1)}\zeta + 8 \pi G \left(- \frac{1}{3} \theta ' p + \frac{1}{2} \rho \:
^{(1)}\zeta\right)\right] = 0, \label{F3} \\[1cm]
&& 2\: ^{(5)}\dot{\tilde{\phi}} +  3 \:^{(4)}\dot{\tilde{\phi}} - 6\: ^{(2)}\phi
^{(2)}\zeta + 8 \pi G \rho \left(\frac{1}{27} {\theta '} ^3 - \frac{1}{2}\:
^{(2)}\zeta\right) = 0, \label{F4} \\[1cm]
&& 6\: ^{(3)}\tilde{\phi} + 2\: ^{(2)}\tilde{\phi} - 16\: ^{(2)}\phi\: ^{(1)}\phi +
16 \pi G (p + 2\rho \:^{(1)}\phi) \nn \\
&& + c^{-2}\left[- ^{(1)}\ddot{\tilde{\phi}} + \frac{11}{2}\: ^{(1)}\zeta^2 -
4\: ^{(2)}\phi\: ^{(1)}\tilde{\phi} + 4 \: ^{(1)}\phi\:^{(2)}\tilde{\phi} + 8
\:^{(1)}\phi \:^{(3)}\tilde{\phi} + 8 \pi G\left(\rho (\;^{(1)}\tilde{\phi}
+ 3\;^{(2)}\phi) - \frac{1}{2}p\;^{(1)}\phi \right)\right] = 0, \label{F5}
\nn\\ \\[1cm]
&& 20\: ^{(5)}\tilde{\phi} + 2\: ^{(4)}\tilde{\phi} + 8\: ^{(2)}\phi ^2 + 32\pi G
\rho \;^{(2)}\phi +  c^{-2}\left[ - ^{(3)}\ddot{\tilde{\phi}} + 11\: ^{(1)}\zeta \:
^{(2)}\zeta - 44 \: ^{(2)}\phi \:
^{(3)}\tilde{\phi} \right. \nn \\
&& \left. - 16\:^{(2)}\phi \:^{(2)}\tilde{\phi} + 32\:^{(1)}\phi
\:^{(5)}\tilde{\phi} - 8\:^{(1)} \phi \:^{(4)}\tilde{\phi} + 48\: ^{(2)}\phi^2 \:^{(1)}\phi + 8\pi G
\left(\rho(^{(3)}\tilde{\phi} + 2\;^{(1)}\phi \frac{1}{9} \theta'^2) -
\frac{1}{2} \:^{(2)}\phi p \right)\right] = 0, \label{F6} \nn \\ \\[1cm]
&& 14\: ^{(4)}\tilde{\phi} - 16 \:^{(2)}\phi ^2 + 16 \pi G \rho \frac{1}{9}
\theta '^2 \nn \\
&& + c^{-2}\left[ - ^{(2)}\ddot{\tilde{\phi}} + 2\: ^{(1)}\zeta \:^{(2)}\zeta
- 64\:^{(2)}\phi^2 \:^{(1)}\phi - 8 \:^{(2)}\phi \:^{(3)}\tilde{\phi} - 16 \:
^{(2)}\phi \:^{(2)}\tilde{\phi} - 16\:^{(1)}\phi \:^{(5)}\tilde{\phi} +
4\:^{(1)}\phi \:^{(4)}\tilde{\phi} \right. \nn \\
&& \left. + 8\pi G \left(2 \frac{1}{9} \theta '^2 (p + 2 \rho \:
^{(1)}\phi) + \rho \:^{(2)}\tilde{\phi}\right)\right] = 0, \label{F7} \\[1cm]
&& ^{(5)}\ddot{\tilde{\phi}}  + \frac{23}{2}\: ^{(2)}\zeta^2 + 60\: ^{(2)}\phi
\:^{(5)}\tilde{\phi} - 20\: ^{(2)}\phi \:^{(4)}\tilde{\phi} - 48\:
^{(2)}\phi\: ^3 - 8\pi  G \rho \left(2\;^{(2)}\phi  \frac{1}{9} \theta '^2 + ^{(5)}\tilde{\phi}\right) = 0, \label{F8} \\[1cm]
&& ^{(4)}\ddot{\tilde{\phi}} - 6 \:^{(2)}\zeta ^2 + 64 \:^{(2)}\phi^3 - 32\:
^{(2)}\phi \:^{(5)}\tilde{\phi} - 12 \:^{(2)}\phi \:^{(4)}\tilde{\phi}\nn \\
&&  - 8 \pi G
\rho \left(4 \:^{(2)}\phi \frac{1}{9} \theta'^2 + 2 \frac{1}{81} \theta'^4 + ^{(4)}\tilde{\phi}\right) = 0. \label{F9}
\end{eqnarray}

The homogeneous harmonic gauge conditions (\ref{hg2}) and (\ref{hg1}) become

\begin{eqnarray}
{^{(1)}\dot{\phi}} & = & - \frac{3}{4} \:^{(1)}\zeta, \label{HG1}\\
{^{(2)}\dot{\phi}} & = & - \frac{5}{4} \:^{(2)}\zeta, \label{HG2}\\
{^{(1)}\dot{\zeta}} & = & 2 \:^{(3)}\tilde{\phi} + 4
\:^{(2)}\tilde{\phi}.\label{HG3} \\
{^{(2)}\dot{\zeta}} & = & 4 \:^{(5)}\tilde{\phi} + 6 \:^{(4)}\tilde{\phi},
\label{HG4} 
\end{eqnarray}

Using (\ref{F3}) and the time dervivative of (\ref{F1}), the harmonic gauge
condition (\ref{HG2}) reads 

\begin{eqnarray}
\dot{\rho} + \theta'(\rho + pc^{-2}) + c^{-2}\left[-\frac{27}{8\pi G}
\:^{(2)}\phi \:^{(1)}\zeta - 4\dot{\rho} \:^{(1)}\phi +  \frac{3}{2} \rho
\:^{(1)}\zeta \right] = 0, \label{F10}
\end{eqnarray}
where (\ref{F2}) and
(\ref{F4}) have been used to express the $c^{-2}$ parts of the equation in this relatively
simple form. This is the continuity equation (\ref{B1}) for the special case of
homogeneity and isotropy.

Next, consider  equations (\ref{F6}) and
(\ref{F7}) with the time derivative of (\ref{F3}). These equations, with the
help of (\ref{F8}) and (\ref{F9}), provide us
with the Euler equation, (\ref{B2i}), for the case of homogeneity and isotropy:

\begin{eqnarray}&& \rho \left(\frac{1}{3} \dot{\theta}' +  \frac{1}{9} \theta'^2 +
2\:^{(2)}\phi\right) + \frac{1}{16\pi G} c^{-2}\left[265
\:^{(2)}\phi^2\;^{(1)}\phi + 132\;^{(2)}\phi\:^{(3)}\tilde{\phi} + 120\;^{(2)}\phi \;^{(2)}\tilde{\phi} \right. \nn \\
&& \left. - 18\:^{(2)}\dot{\phi} \;^{(1)}\zeta - 18\;^{(2)}\phi
\;^{(1)}\dot{\zeta} - 30\;^{(1)}\zeta\;^{(2)}\zeta + 80 \;^{(1)}\phi
^{(5)}\tilde{\phi} + 120 \;^{(1)}\phi\;^{(4)}\tilde{\phi} + 54\;
^{(2)}\phi\;^{(1)}\zeta \frac{1}{3} \theta' \right. \nn \\
&& \left. + 16\pi G \left(- \dot{\rho} \left(\frac{1}{2}\;^{(1)}\zeta +
\;^{(1)}\phi \frac{1}{3}\theta'\right) + \dot{p} \frac{1}{3}\theta' + p 
\left( \frac{1}{3}\dot{\theta'} + \frac{1}{2}\;^{(2)}\phi
- 7 \; \frac{1}{9}\theta'^2\right) \right. \right. \nn \\
&& \left. \left. + \rho \left(- \frac{1}{2}\;^{(1)}\dot{\zeta} -
^{(3)}\tilde{\phi} - 14 \;^{(1)}\phi \;\frac{1}{9}\theta'^2 -
2\;^{(2)}\tilde{\phi} - 24\;^{(2)}\phi\;^{(1)}\phi - 8\;^{(1)}\dot{\phi}
\;\frac{1}{3} \theta + \frac{1}{2}\; ^{(1)}\zeta \theta'\right)\right)\right]
= 0.\nn\\
\label{F11}
\end{eqnarray}

We may now read off the potentials $\phi$, $\zeta_i$ and $\phi_{ij}$ from the FRW metric
(\ref{metric}). With the help of the harmonic gauge conditions (\ref{HG1}) to
(\ref{HG4}) we are able to write down the functions $^{(1)}\phi$,
$^{(2)}\phi$ etc. which appear in the ans\"atze (\ref{ant}):
	
\begin{eqnarray}
^{(1)}\phi & = & \dot{A},  \nn \\
^{(2)}\phi & = & - \frac{1}{2} \frac{\ddot{R}}{R},  \nn \\
^{(1)}\zeta  & = &  - \frac{4}{3} \ddot{A},  \nn \\
^{(2)}\zeta  & = &   \frac{2}{5}\left(\frac{\dddot{R}}{R} - \frac{\dot{R}\ddot{R}}{R^2}\right),  \nn \\
^{(1)}\tilde{\phi}  & = &   4 \dot{A}^2,  \nn \\
^{(2)}\tilde{\phi}  & = &   - \frac{2}{5} \dddot{A} + \frac{8}{15} \ddot{A}
\frac{\dot{R}}{R} + \frac{2}{5} \dot{A} \frac{\ddot{R}}{R} + 2 \dot{A} \left(\frac{\dot{R}}{R}\right)^2,  \nn \\
^{(3)}\tilde{\phi}  & = &  \frac{2}{15} \dddot{A} - \frac{16}{15} \ddot{A}
\frac{\dot{R}}{R} - \frac{4}{5} \dot{A} \frac{\ddot{R}}{R} - 4 \dot{A} \left(\frac{\dot{R}}{R}\right)^2,  \nn \\
^{(4)}\tilde{\phi}  & = &  \frac{1}{21} \frac{\ddddot{R}}{R} + \frac{9}{35} \frac{\dddot{R}}{R} \frac{\dot{R}}{R} - \frac{16}{21} \left(\frac{\ddot{R}}{R}\right)^2 + \frac{29}{28} \left(\frac{\dot{R}}{R}\right)^4 - \frac{39}{35} \frac{\ddot{R}}{R} \left(\frac{\dot{R}}{R}\right)^2,  \nn \\
^{(5)}\tilde{\phi}  & = &   \frac{4}{35} \frac{\ddddot{R}}{R} - \frac{87}{14}
\left(\frac{\dot{R}}{R}\right)^4 + \frac{219}{210}
\left(\frac{\ddot{R}}{R}\right)^2 - \frac{32}{105} \frac{\dddot{R}}{R}
\frac{\dot{R}}{R} + \frac{92}{105} \frac{ \ddot{R}}{R}
\left(\frac{\dot{R}}{R}\right)^2.  \nn \\
\label{Fullpot}
\end{eqnarray}
	
Using these potentials in (\ref{F11}) and adopting a power expansion for
$\theta'$  \\

\hspace {2cm} $\theta' = \theta + c^{-2} \theta'',$ \hspace{2cm} where $\theta = 3
\frac{\dot{R}}{R}$, \\

one obtains, with the help of (\ref{F5}) and after considerable simplification
	
\begin{eqnarray}
3 \frac{\ddot{R}}{R} & = & - 4 \pi G (\rho + 3pc^{-2}) + c^{-2}\left[-
\dddot{A} + 4 \ddot{A} \frac{\dot{R}}{R} - 3 \dot{A} \frac{\ddot{R}}{R}\right].
\label{FullRay}
\end{eqnarray}
        
This is the Raychadhuri equation of general relativity (\ref{Ray}), with the
required inclusion of the pressure term, which was
missing in the Newtonian case (\ref{nop}), and with $c^{-2}$
corrections. Since $\dot{A} \ll c^{2}$, we can be sure that the corrections
are of higher order and do not contribute to the theory at the $c^0$ level. We may now use this result, along with the now known values for the potentials (\ref{Fullpot}), in (\ref{F10}) to derive the Friedmann equation
	
\begin{eqnarray}
\left(\frac{\dot{R}}{R}\right)^2 = \frac{8 \pi G}{3} \rho + \Upsilon c^{-2},
\label{FullFried}
\end{eqnarray}
where $\Upsilon$ is defined to be the solution of the differential equation 

\begin{eqnarray}
\dot{\Upsilon} + 2 \frac{\dot{R}}{R} \Upsilon + \frac{2}{3} \dddot{A}
\frac{\dot{R}}{R} + \ddot{A}\left( - \frac{2}{3}
\left(\frac{\dot{R}}{R}\right)^2 + 6 \frac{\ddot{R}}{R}\right) +  \dot{A}
\left( 5 \frac{\ddot{R}\dot{R}}{R^2} - 3 \left(\frac{\dot{R}}{R}\right)^3\right) +
 2 \theta'' \left(\frac{1}{3} \frac{\ddot{R}}{R} - 3
\left(\frac{\dot{R}}{R}\right)^2\right) = 0. \nn
\end{eqnarray}

The Bianchi identities (\ref{F10}) and
(\ref{F11}), the field equations (\ref{F1}) to (\ref{F9}) and the harmonic gauge
conditions (\ref{HG1}) to (\ref{HG4}) may be understood in the following
manner: The time derivative of equation (\ref{F3}) along with the equations
(\ref{F6}), (\ref{F7}), (\ref{F8}), (\ref{F9}) and with the aid of the harmonic gauge conditions give us the
Bianchi identity (\ref{F11}) which leads to the Raychadhuri equation (\ref{FullRay}). The time derivative of equation (\ref{F1}) with equations
(\ref{F3}), (\ref{F2}), (\ref{F4}) and the harmonic gauge condition (\ref{HG2}) gives the Bianchi
identity (\ref{F10}), which leads to the Friedmann equation (\ref{FullFried}). This leaves us
with the field equation (\ref{F5}) which may also be combined with (\ref{F1})
to give the Raychadhuri equation. 

It may also be shown that the following relationships between the field
equations exist. The time derivative of (\ref{F1}) is (\ref{F3}).    The time
derivative of (\ref{F2}) with the aid of (\ref{F5}) is (\ref{F4}). Equations
(\ref{F6}) and (\ref{F7}) may be combined in such a way as to be (\ref{F1}). Finally,   Equations
(\ref{F8}) and (\ref{F9}) may be combined in such a way as to be the second time
derivative of (\ref{F4}). 

Thus the post-Newtonian approximation may be completely defined for the FRW
cosmology with 
the Raychadhuri equation (\ref{FullRay}), the Friedmann equation (\ref{FullFried}) and an
extended Poisson equation (\ref{F1}), which after substitution of
(\ref{Fullpot}), becomes

\begin{eqnarray}
3\frac{\ddot{R}}{R} = - 4\pi G\rho + c^{-2}(-\dddot{A} - 16 \pi G \rho \dot{A}).
\label{FullA}
\end{eqnarray}

The unknown $\theta''$ may be set to zero without loss of
generality. Equation (\ref{FullFried}) may be equivalently represented by
(\ref{F10}). Given an equation of state ($p
c^{-2} =
w \rho$, $w$ an arbitrary constant), the system (\ref{F10}), (\ref{FullRay})
and (\ref{FullA}) gives equations for $\dot{\rho}$, $\ddot{R}$ and
$\dddot{A}$ respectively, and thus, forms a well-posed set, allowing the unknowns $R(t)$, $\rho (t)$ and $A(t)$ to be
determined.

It is the gravitational potential $A(t)$ that incorporates the pressure into the
theory. The pressure and $A(t)$ are related through a  third order
differential equation one can obtain by combining (\ref{FullRay}) and
(\ref{FullA}). Varying the equation of state therefore results in varying
solutions for the density $\rho(t)$ and for $R(t)$, which are analogous to
those solutions predicted by the fully general relativistic theory for the
FRW cosmologies, (\ref{sol}). Setting $A = 0$, for example, implies vanishing pressure, $p
= 0$, which is the post-Newtonian approximation of FRW for dust. In this case
the higher order corrections of the post-Newtonian theory disapear and one
ends up with the Newtonian theory, which we have shown in Section \ref{n}
describes the case of dust.

\section{Concluding Remarks}
\label{conc}

The FRW
cosmology is a very good approximation to the large scale structure of the
universe, at least to the present epoch \cite{hawk}. Due to its obvious
simplicity, Newtonian approximations, where they reproduce results which are similar to
the fully general relativistic theory, are preferable. General relativity has a well-posed Cauchy problem in the case of perfect
fluids with a barotropic equation of state \cite{synge}. Newtonian theory, on the other
hand, is not well-posed \cite{szek-rain}. We have seen that the Newtonian theory reproduces the results of it's general
relativistic counterpart only for the special case of dust, ie. for a matter-dominated universe with an equation of state, $p =
\rho c^2$. Although,  any equation of state may be written down, it won't
make any difference since the pressure
does not appear in the dynamics.

The post-Newtonian
theory, which is consistent and well-posed, does provide field equations with
 pressure entering the dynamics through the potential $A(t)$. We are thus able to vary the equation of
state and, in doing so, will obtain various solutions for R(t) and the
density, $\rho$. Hence, the post-Newtonian theory seems to be a favourable
approximation of the
fully general relativistic theory.

I would like to acknowledge Peter Szekeres for useful discussions.

\end{document}